\pdfoutput=1
\documentclass[sigconf]{acmart}
\AtBeginDocument{%
  }

\copyrightyear{2027}
\acmYear{2027}
\setcopyright{cc}
\setcctype{by}
\acmConference[ASPDAC '27]{32nd Asia and South Pacific Design Automation Conference}{January 25--28, 2027}{Tokyo, Japan}
\acmBooktitle{32nd Asia and South Pacific Design Automation Conference (ASPDAC '27), January 25--28, 2027, Tokyo, Japan}
\acmDOI{10.1145/3840404.3848160}
\acmISBN{979-8-4007-2923-2/2027/01}

\usepackage{tikz}
\usetikzlibrary{shapes.geometric, arrows.meta, positioning}
\usepackage[ruled,vlined,linesnumbered]{algorithm2e}
\usepackage{multirow}
\begin{document}

\title[3D-VASE: Volumetric-Aware Scalable and Efficient Multi-Layer Design Synthesis for 3D-Printed Microfluidics]{3D-VASE: \underline{V}olumetric-\underline{A}ware \underline{S}calable and \underline{E}fficient Multi-Layer Design Synthesis for \underline{3D}-Printed Microfluidics}

%\renewcommand{\shortauthors}{Anonymous et al.}


% \author[Zhang et al.]{Yushen Zhang$^{*}$, Jiashuo Chen$^{*}$, Tsun-Ming Tseng, Ulf Schlichtmann}
% \affiliation{%
%   \institution{{\textit{Technical University of Munich}}, Germany}
%   \city{\{yushen.zhang, jiashuo.chen, tsun-ming.tseng, ulf.schlichtmann\}@tum.de\\}
%   \country{{\footnotesize $^{*}$These authors contributed equally.}}
% }

\author{Yushen Zhang}
\authornote{These authors contributed equally.}
\affiliation{%
  \institution{Technical University of Munich}
  \city{Munich}
  \country{Germany}}
\email{yushen.zhang@tum.de}

\author{Jiashuo Chen}
\authornotemark[1]
\affiliation{%
  \institution{Technical University of Munich}
  \city{Munich}
  \country{Germany}}
\email{jiashuo.chen@tum.de}

\author{Tsun-Ming Tseng}
\affiliation{%
  \institution{Technical University of Munich}
  \city{Munich}
  \country{Germany}}
\email{tsun-ming.tseng@tum.de}

\author{Ulf Schlichtmann}
\affiliation{%
  \institution{Technical University of Munich}
  \city{Munich}
  \country{Germany}}
\email{ulf.schlichtmann@tum.de}


\begin{abstract}
Additive manufacturing (3D printing) enables complex multi-layer microfluidic devices, but the physical synthesis of passive-flow designs remains a key bottleneck. In such systems, designers must encode timing synchronization and target dosing directly into channel geometry, even though the available routing space across multiple 3D layers is often highly congested, which makes scaling to complex designs difficult. Existing approaches formulate placement, routing, and fluidic constraints as a single monolithic optimization, leading to combinatorial explosion and poor performance on complex designs.
This paper presents a volumetric-aware two-stage synthesis framework that decouples global structure from routing detail. A Mixed-Integer Programming (MIP) model first computes a compact multi-layer placement, followed by a dynamic 3D routing engine that enforces timing and volumetric constraints during path construction. To further improve routability in dense layouts, we introduce a hierarchical conflict resolution strategy combining local rip-up-and-reroute with controlled global expansion.
Experimental results show that, under timing and volumetric constraints, 3D-VASE consistently improves synthesis speed and, in most test cases, also improves layout quality. Notably, for difficult large-scale cases where prior methods often failed to return any solution within the time limit, 3D-VASE produces feasible high-quality layouts within practical runtimes. These advances enable fast, iterative design-space exploration for complex 3D-printed microfluidic systems.
\end{abstract}




\maketitle
\section{Introduction}
\begin{figure}[t]
    \centering
    \includegraphics[width=\linewidth]{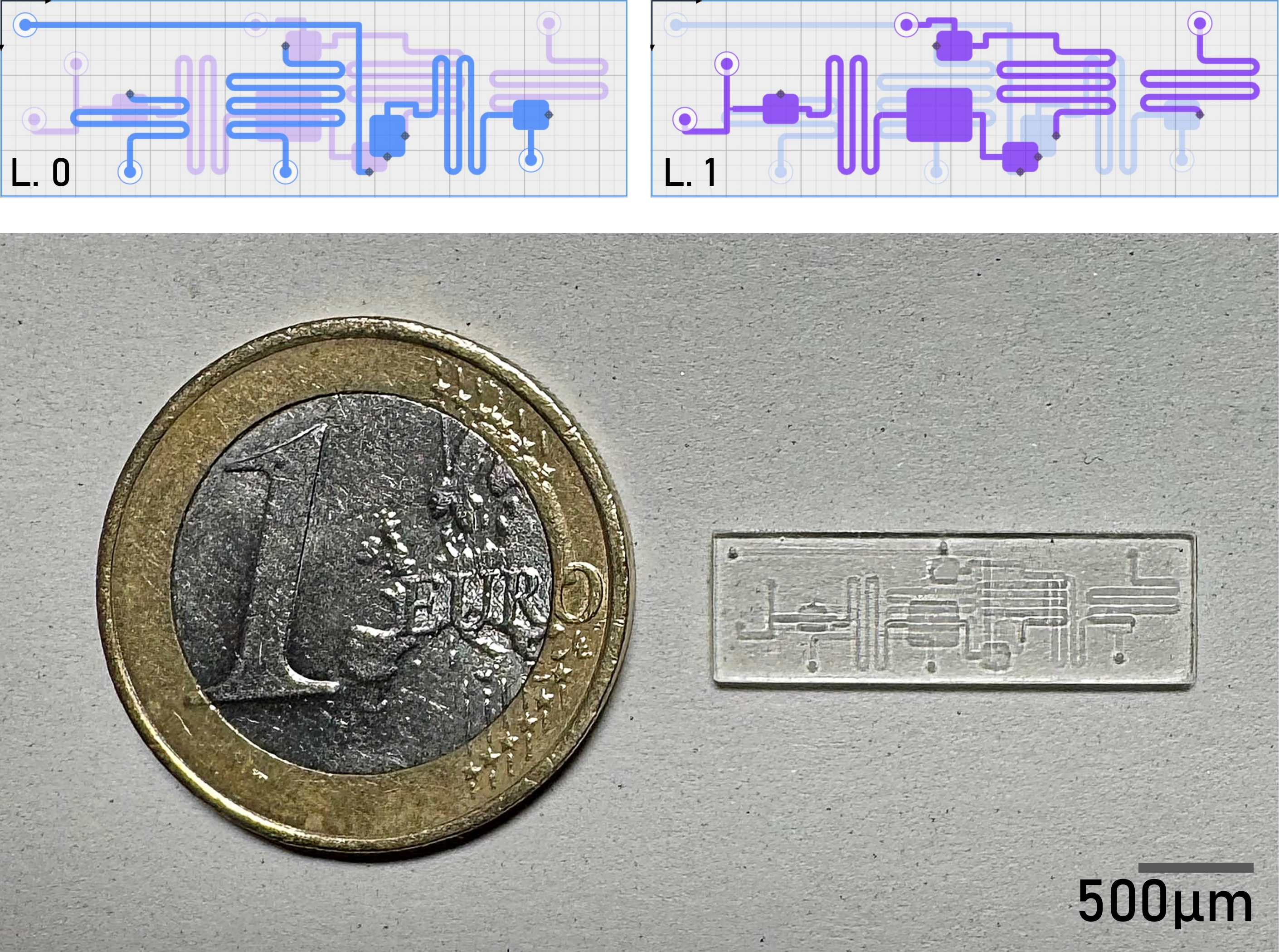}
 % \vspace{-0.6cm}
    \caption{Design and fabrication of a 3D-VASE synthesized 3D-printed microfluidic chip. (Top) Layout of the microfluidic channel network showing two design layers. (Bottom) Optical image of the fabricated chip alongside a €1 coin for scale.}
    \label{fig:fab-example}

    % \vspace{-0.4cm}
\end{figure}

Microfluidic devices enable miniaturized and automated laboratory processes in applications ranging from diagnostics and drug delivery to cell biology and chemical synthesis \cite{whitesides2006origins,dittrich2006lab,sackmann2014present}. Conventional fabrication techniques such as photolithography and polydimethylsiloxane (PDMS) glass bonding have enabled many advances, but they rely on complex cleanroom workflows and are largely limited to planar layouts \cite{halldorsson2015advantages,sia2003pdms}. Three-dimensional printing (3D printing), especially stereolithography (SLA), offers rapid prototyping, reduced fabrication barriers, and vertically stacked flow layers with internal channels and vias \cite{au20163d,nielsen20203d,bhattacharjee2016upcoming,waheed2016enablers,shallan2014cost,comina2014pdms}. Figure \ref{fig:fab-example} shows such a microfluidic chip printed with an SLA 3D printer.
However, the structural freedom of multi-layer 3D printing also makes device design more difficult. Designers must coordinate fluid dynamics, channel geometry, layer organization, manufacturability, and assay requirements in a three-dimensional space, often through manual CAD modeling and trial-and-error refinement. This process becomes slow and error-prone as chip complexity increases.

In multi-layer 3D-printed microfluidic chips, fluid control is fundamentally a geometric design problem. Many assays require strict synchronization so that multiple streams arrive at a reaction chamber simultaneously and with prescribed mixing ratios \cite{song2003microfluidic,ward2015mixing}. Compared with valve-rich platforms, such as conventional PDMS chips that rely on active on-chip actuation for timing and dosing, many 3D-printed systems primarily use passive flow layers, meaning that fluid behavior must be programmed directly into channel geometry \cite{halldorsson2015advantages}. In this setting, geometry is not merely a packaging concern; it becomes the mechanism for achieving functional correctness.

In passive-flow systems, the arrival time $T$ is fundamentally determined by channel hold-up volume $V$ and inlet flow rate $Q$: when $Q$ is fixed by the pump, a larger $V$ leads to later arrival. Consequently, channel length, cross-sectional area, layer transitions, and routing detours directly influence both arrival timing and delivered volume. This dependence makes the design space extremely sensitive to geometric changes in the routing layout. A modification that appears minor from a geometric perspective, such as shifting a path to avoid congestion or inserting a detour around an obstacle, can change $V$ enough to disturb synchronization or dosage targets. As a result, manual design refinement becomes slow and error-prone, especially in dense multi-layer layouts where spatial constraints and fluidic constraints interact continuously.

Translating the above fluidic mechanism into a design automation problem is highly challenging. Prior microfluidic design automation has largely focused on electrode-array-driven digital microfluidics \cite{fair2007digital,su2008high,xu2008defect,cho2003droplet}, planar or interactive continuous-flow CAD support \cite{sanka20193d}, control-layer-based large-scale integration \cite{tseng2019cloud}, or manual multi-layer modeling for 3D-printed devices \cite{zhang2024open}. These tools improve design productivity, but they do not provide an end-to-end automatic placement-and-routing flow for passive-flow 3D-printed chips under timing and volumetric constraints. The closest prior synthesis framework, 3M-DeSyn, formulates placement, routing, and fluidic constraints in a single monolithic optimization model \cite{zhang20253m}. While this preserves strong constraint fidelity, its tightly coupled formulation suffers from severe combinatorial explosion as netlist size grows and cross-layer detours become more frequent.

To break this scalability barrier, this paper proposes a novel, two-stage physical synthesis framework that fundamentally decouples global placement from detailed routing. The primary contribution of this work lies in replacing the monolithic mathematical model with a pipeline tailored for multi-layer 3D-printed microfluidics. Specifically, we introduce a Mixed-Integer Programming (MIP)-based placement estimation model paired with a dynamic 3D routing verification algorithm, effectively mapping fluidic timing and volumetric constraints into geometric parameters. Furthermore, we develop a hierarchical conflict resolution mechanism to dynamically mitigate routing congestion in high-density layouts. Extensive evaluations on multiple test cases demonstrate substantial speedups over state-of-the-art methods and successful routing on complex, large-scale designs where existing solvers fail, providing a robust automated solution for fast design iteration and complex multi-layer layout generation. 

\section{Problem Formulation}
We formulate physical synthesis as automated placement and routing for a multi-layer 3D-printed microfluidic chip. A chip contains modules and connecting channels distributed across stacked physical layers, with vertical vias enabling inter-layer flow.

\subsection{Input and Output}
The input is a microfluidic netlist containing modules $M$, directed fluidic connections $C$, optional inlet flow rates $Q$, manufacturing parameters, and optional timing/volumetric constraints. Each module $m\in M$ has a fixed footprint $(w_m,l_m)$ on an assigned layer, and each connection $c\in C$ specifies its source, target, channel width $w_c$, and height $h_c$. Manufacturing parameters include the maximum number of layers, layer height, minimum clearance $\rho$, and possible layer-assignment restrictions. The output is a print-ready 3D layout, including the coordinate and layer $(x,y,ly)$ of each module and a collision-free 3D geometric path for every connection.

\subsection{Timing and Volumetric Constraints}

Unlike active-valve platforms, passive 3D-printed microfluidic chips encode timing and dosing directly in geometry. A routed channel determines its hold-up volume through its path length, cross-sectional area, and layer transitions; therefore, even a small detour inserted for obstacle avoidance can change arrival time or delivered dosage. In our formulation, timing and volumetric constraints are not independent requirements but are coupled by the relation between arrival time $T$, channel hold-up volume $V$, and inlet flow rate $Q$:
\begin{equation}
T = V / Q
\end{equation}
A timing constraint $T_r$ requires converging streams to reach a target mixer simultaneously. For two paths $A$ and $B$, this requires
\begin{equation}
\frac{V_A}{Q_A} = \frac{V_B}{Q_B}
\end{equation}
A volumetric constraint $V_r$ specifies the target mixing ratio through the assigned inlet flow-rate ratio. To avoid initial fluid waste, these streams must also satisfy simultaneous arrival under the same flow-rate relation. Thus, both timing synchronization and ratio-based metering reduce to the unified T/V consistency condition
\begin{equation}\label{eq:tv_mismatch}
|V_A \cdot Q_B - V_B \cdot Q_A| \le \epsilon
\end{equation}
where $V_A,V_B$ are geometric hold-up volumes, $Q_A,Q_B$ are inlet flow rates, and $\epsilon$ covers grid quantization and numerical tolerance. This formulation allows both synchronization and ratio-based metering to be handled by geometric path design. If a deliberate timing offset is required, the same constraint can be extended by adding the corresponding target volume offset to the relevant path.

\subsection{Physical Design Objectives}
The synthesis process optimizes compactness and routing efficiency. We use the chip bounding-box objective $F_{size}$ and total routing objective $F_{route}$:
\begin{equation}
\min F_{size} = \alpha \cdot w_{chip} + \beta \cdot l_{chip} + \gamma \cdot nl_{chip}
\end{equation}
\begin{equation}
\min F_{route} = \sum_{c \in C} L(c)
\end{equation}
where $w_{chip}$, $l_{chip}$, and $nl_{chip}$ denote chip width, chip length, and layer count, and $L(c)$ is the routed length of channel $c$. 3D-VASE minimizes these metrics cooperatively across placement and routing rather than solving a monolithic Pareto optimization.

\section{Methodology}
To overcome the computational scalability bottleneck of traditional single-step global optimization models in multi-layer physical spaces, 3D-VASE adopts a strict dimensionality-reduction and decoupling architecture. We partition the complex physical synthesis problem into two independently computed stages with unidirectional data transfer: a Mixed-Integer Programming (MIP)-based global placement stage and a heuristic-search-based 3D routing stage. 

\begin{figure}[t]
\centering
\resizebox{0.98\columnwidth}{!}{
\begin{tikzpicture}[
    node distance=0.58cm,
    base/.style={draw, rounded corners=2.6pt, align=center, font=\normalsize, minimum height=1.06cm, inner sep=4pt},
    io/.style={base, trapezium, trapezium left angle=70, trapezium right angle=110, minimum width=3.45cm, text width=3.22cm, fill=blue!5, draw=blue!55!black},
    process/.style={base, rectangle, minimum width=3.45cm, text width=3.22cm, fill=teal!6, draw=teal!55!black},
    feedbackbox/.style={base, rectangle, minimum width=3.15cm, text width=2.92cm, fill=orange!8, draw=orange!65!black},
    decision/.style={base, diamond, aspect=1.7, minimum width=3.05cm, text width=2.35cm, inner sep=1.2pt, fill=purple!7, draw=purple!60!black},
    arrow/.style={line width=0.88pt, ->, >=Stealth}
]

\node (input) [io, minimum width=2.2cm, text width=1.9cm]{Input Netlist};
\node (stage1) [process, below=0.62cm of input] {Stage 1: MIP-Based\\Global Placement};
\node (stage2) [process, below=0.62cm of stage1] {Stage 2: Search-Based\\3D Routing};
\node (check_conflict) [decision, below=0.58cm of stage2] {Routing\\Conflict?};
\node (check_volume) [decision, below=0.58cm of check_conflict] {Volume\\Met?};
\node (output) [io, below=0.65cm of check_volume] {Print-Ready 3D\\Modeling File};

\node (feedback) [feedbackbox, right=1.1cm of stage1] {Global Feedback\\(Relax Constraints)};
\node (ripup) [feedbackbox] at (feedback |- check_conflict) {Local Rip-Up \&\\Reroute};
\node (detour) [feedbackbox, left=1.1cm of check_volume] {Detour\\(Volume Adjustment)};

\draw [arrow] (input) -- (stage1);
\draw [arrow] (stage1) -- (stage2);
\draw [arrow] (stage2) -- (check_conflict);
\draw [arrow] (check_conflict) -- node[left, font=\normalsize] {No} (check_volume);
\draw [arrow] (check_conflict) -- node[above, font=\normalsize] {Yes} (ripup.west);
\draw [arrow] (check_volume) -- node[left, font=\normalsize] {Yes} (output);
\draw [arrow] (check_volume) -- node[above, font=\normalsize] {No} (detour.east);
\draw [arrow] (detour.north) |- node[pos=0.74, above, font=\normalsize] {Retry} (stage2.west);
\draw [arrow] ([xshift=-0.62cm]ripup.north) |- node[pos=0.74, above, font=\normalsize] {Success} (stage2.east);
\draw [arrow] ([xshift=0.62cm]ripup.north) -- node[right, font=\normalsize] {Fail} ([xshift=0.62cm]feedback.south);
\draw [arrow] (feedback) -- (stage1);

\end{tikzpicture}
}

\caption{Flow of the 3D-VASE two-stage synthesis framework with hierarchical conflict resolution.}
\Description{Flow of the two-stage synthesis pipeline. Input netlist enters MIP-based global placement, then search-based 3D routing. Routing conflicts trigger local rip-up and reroute; if no conflict but target volume is not met, detour adjustment is applied and routing retries. Repeated rip-up failure triggers global feedback that relaxes placement constraints and restarts Stage 1.}
\label{fig:flowchart}
% \vspace{-0.3cm}
\end{figure}
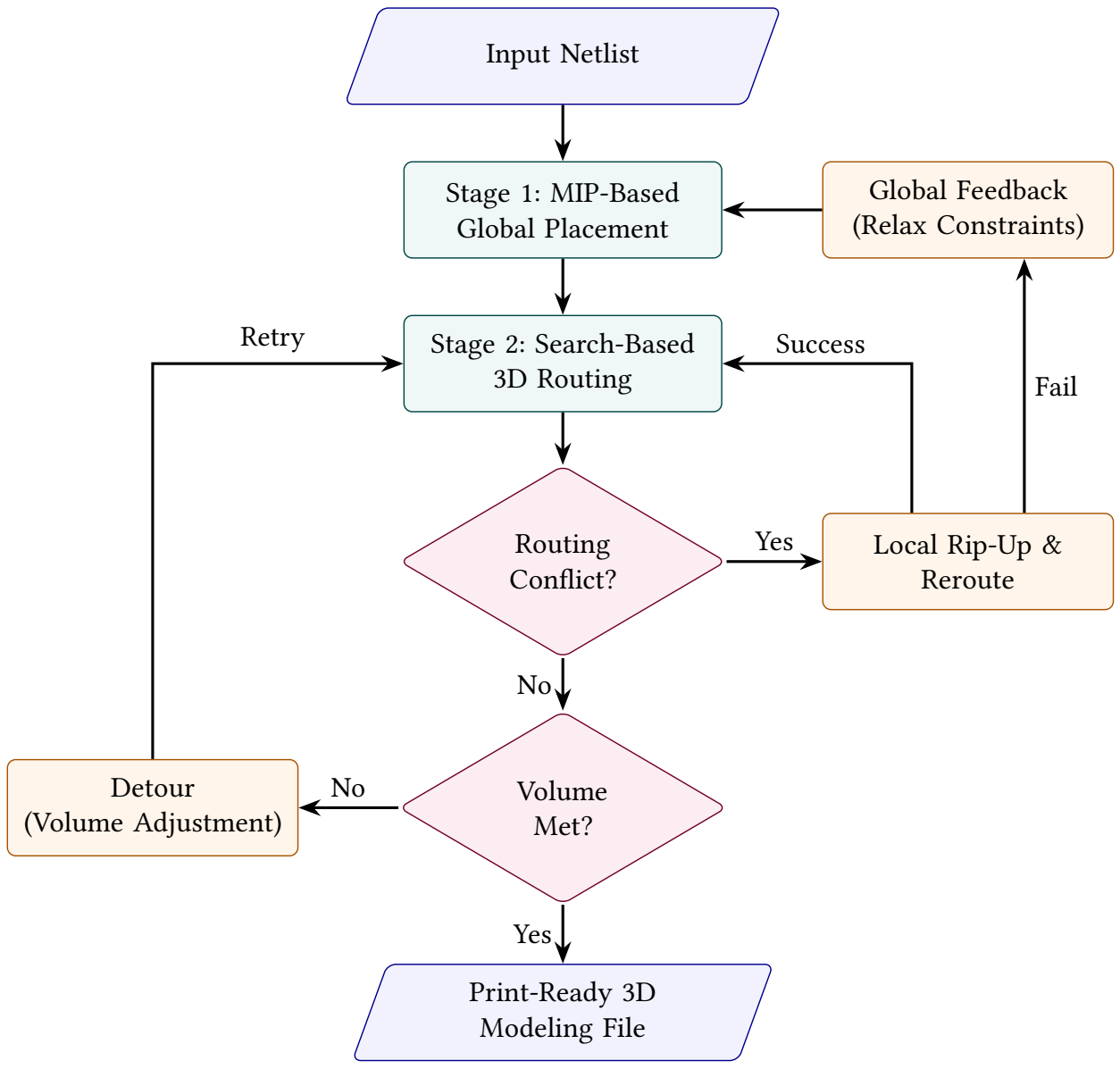

Figure~\ref{fig:flowchart} summarizes the end-to-end 3D-VASE workflow and clarifies the branching between conflict handling (rip-up/reroute) and volume correction (detour).

\subsection{Two-Stage Physical Synthesis Framework}
\subsubsection{\textbf{Stage 1: MIP-based Global Placement}}
The core task of the first stage is to determine the 2D placement of all modules on their assigned layers while compressing the overall chip extent. We model each physical layer as a continuous 2D bounding-box packing space and establish inter-layer relationships through integer variables.
For any module $m \in M$, its physical state is defined by the top-left coordinate $(x_m, y_m)$, layer index $ly_m$, bottom-right coordinate $(x2_m, y2_m)$, and four binary orientation variables $d_{0}, d_{90}, d_{180}, d_{270} \in \{0, 1\}$. Here, $w_m$ and $l_m$ are the fixed width and length of module $m$. The bottom-right coordinate is computed by
\begin{equation}
x2_m = x_m + (d_0 + d_{180}) \cdot w_m + (d_{90} + d_{270}) \cdot l_m
\end{equation}
\begin{equation}
y2_m = y_m + (d_0 + d_{180}) \cdot l_m + (d_{90} + d_{270}) \cdot w_m
\end{equation}
A mutual exclusion constraint must be simultaneously applied to ensure the module occupies only a single physical orientation:
\begin{equation}
d_0 + d_{90} + d_{180} + d_{270} = 1
\end{equation}

To make placement estimation physically consistent with actual channel endpoints, we compute each pin location from module placement, orientation, and pin offset ratios. For a pin $p$ on module $m$, let $(r_x^{(p)}, r_y^{(p)}) \in [0,1]^2$ denote its normalized offset ratios in the module's local frame. Its absolute coordinates are estimated as
\begin{equation}
\begin{aligned}
x_{pin}(m,p) =\;& x_m + d_0\, r_x^{(p)} w_m + d_{180}(1-r_x^{(p)})w_m \\
& + d_{90}(1-r_y^{(p)})l_m + d_{270} r_y^{(p)} l_m.
\end{aligned}
\end{equation}
\begin{equation}
\begin{aligned}
y_{pin}(m,p) =\;& y_m + d_0\, r_y^{(p)} l_m + d_{180}(1-r_y^{(p)})l_m \\
& + d_{90} r_x^{(p)} w_m + d_{270}(1-r_x^{(p)})w_m.
\end{aligned}
\end{equation}
For a connection $c=(p_s,p_t)$ between source pin $p_s$ and target pin $p_t$, where $m_s$ and $m_t$ denote the modules containing $p_s$ and $p_t$, we estimate routing demand by pin-to-pin Manhattan distance:
\begin{equation}
\begin{aligned}
L_{est}(c)=\;&\left|x_{pin}(m_s,p_s)-x_{pin}(m_t,p_t)\right| \\
&+ \left|y_{pin}(m_s,p_s)-y_{pin}(m_t,p_t)\right|.
\end{aligned}
\end{equation}

Objective Function: We minimize chip footprint and estimated channel length with a linear weighted objective:
\begin{equation}
\min \; \alpha \cdot w_{chip} + \beta \cdot l_{chip} + \gamma \cdot nl_{chip} + \omega \sum_{c \in C} L_{est}(c)
\end{equation}
where $w_{chip}$, $l_{chip}$, and $nl_{chip}$ are chip width, chip length, and layer count.
$L_{est}(c)$ is the Manhattan estimate of channel length for connection $c$.
The coefficients $\alpha, \beta, \gamma, \omega$ are user-defined scalar weights.

Non-overlapping Constraints: To prevent module overlap, any two modules $m_a$ and $m_b$ must be separated in at least one of six orthogonal directions (left, right, front, back, up, down). We linearize this disjunction with a Big-M model using large constant $M$ and binary variables $q_1, \ldots, q_6 \in \{0,1\}$:
\begin{subequations}
\begin{align}
x2_{m_a} + \rho &\le x_{m_b} + q_1 \cdot M \\
x2_{m_b} + \rho &\le x_{m_a} + q_2 \cdot M \\
y2_{m_a} + \rho &\le y_{m_b} + q_3 \cdot M \\
y2_{m_b} + \rho &\le y_{m_a} + q_4 \cdot M \\
ly_{m_a} + 1 &\le ly_{m_b} + q_5 \cdot M \\
ly_{m_b} + 1 &\le ly_{m_a} + q_6 \cdot M
\end{align}
\end{subequations}
where $\rho$ is the manufacturable minimum clearance. We enforce
\begin{equation}
\sum_{i=1}^{6} q_i = 5,
\end{equation}
so exactly one inequality is active ($q_i=0$), which guarantees separation in at least one direction.
\subsubsection{\textbf{Stage 2: Search-based 3D Routing}}
After completing the MIP solving, the fixed module coordinates and layer indices are extracted and input into the second stage. We discretize the internal unoccupied physical volume of the entire chip into a 3D Bounded-Sliceline Grid (BSG) and strictly define it mathematically as a directed graph $G=(V, E)$. Algorithm \ref{alg:routing_search} outlines this.

Vertex Set $V$ represents the discrete coordinate points $(x, y, z)$ in the physical space. The physical spacing between adjacent vertices is strictly equal to the minimum effective print resolution $\mu$ of the 3D printer. For all physical coordinates falling within the 3D bounding boxes of the modules, their corresponding vertices $v \in V$ are directly assigned an infinite cost value and are permanently marked as non-traversable static obstacles in graph $G$.

Edge Set $E$ represents the physical connectivity for fluid flow between adjacent spatial coordinates. Each vertex $v$ has up to 6 outgoing edges (corresponding to the six orthogonal directions in 3D space). The routing cost is accumulated step by step as
\begin{equation}
g(n) = g(parent) + Cost_{move} + Penalty_{bend} + Penalty_{via}.
\end{equation}
Here, $Cost_{move}$ is the unit step cost for moving to an adjacent grid vertex, $Penalty_{bend}$ is added when the moving direction changes between two consecutive steps, and $Penalty_{via}$ is added when $z$ changes and a vertical via is created. In physical terms, these three components respectively capture basic path length, bend-induced hydraulic penalty, and cross-layer transition overhead. This formulation therefore discourages unnecessary detours, bends, and vias, which otherwise increase hydraulic loss and dead volume.

Based on this acummulated routing cost, we run a 3D A* search on graph $G$. The node priority is
\begin{equation}
f(n) = g(n) + h(n),
\end{equation}
where $h(n)$ is the Manhattan distance from node $n$ to the target. Thus, A* combines the realized physical cost $g(n)$ with a heuristic estimate of the remaining distance, enabling efficient search while preserving the routing preferences.

\begin{algorithm}
\caption{Volume-Constrained 3D A* Routing}
\label{alg:routing_search}
\KwIn{Start node $s$, End node $e$, Target volume $V_{req}$, BSG space $G$, Volume tolerance $\delta$}
\KwOut{Routed path $P$, or empty set $\emptyset$}
\tcp{Notation: $Open/Closed$ are A* frontier/visited sets; $Dist=\mu$ is one grid step}
\tcp{$P_{bend},P_{via}$ are bend/via penalties; $A_{ch}=w_c h_c$ is channel area}
\tcp{$n.vol$ is accumulated volume from $s$ to node $n$}

$Open \leftarrow \{s\}$\;
$Closed \leftarrow \emptyset$\;

\While{$Open \neq \emptyset$}{
    $n \leftarrow Open.\text{popNodeWithLowestFCost}()$\;
    $Closed \leftarrow Closed \cup \{n\}$\;
    
    \If{$n.pos == e.pos$}{
        \eIf{$|n.vol - V_{req}| \le \delta$}{
            \Return $\text{ReconstructPath}(n)$\;
        }{
            \textbf{continue} \tcp*{Reject endpoint state if volume is outside tolerance}
        }
    }
    
\ForEach{adjacent node $n_{next}$ of $n$}{
        \If{$n_{next}$ is valid and not occupied in $G$ \textbf{and} $n_{next} \notin Closed$}{
            $cost \leftarrow Dist + P_{bend} + P_{via}$\;
            $n_{next}.vol \leftarrow n.vol + Dist \times A_{ch}$\;
            $n_{next}.gCost \leftarrow n.gCost + cost$\;
            Update $Open$ with $n_{next}$\;
        }
    }
}
\Return $\emptyset$\;
\end{algorithm}
% \vspace{-0.5cm}

\subsection{Timing and Volumetric Rigid Constraint Mechanism}
As defined in Section 3.1, 3D-VASE maps timing and volumetric requirements into the unified T/V consistency constraint in Eq.~\ref{eq:tv_mismatch}. The framework enforces this condition in both placement and routing.

\subsubsection{\textbf{Volume Pre-evaluation in the Placement Stage}}
To prevent topological deadlocks, 3D-VASE proactively enforces volumetric feasibility during global placement. Since the actual 3D fluidic paths are not yet routed, we estimate the channel volume from the placement-stage Manhattan length $L_{\mathrm{est}}$. Let $\eta_s$ denote the empirical detour scaling factor that compensates for the fact that the final routed path is typically longer than the pure Manhattan estimate. The expected channel volume is then
\begin{equation}
V_{\mathrm{est}} = \eta_s \, L_{\mathrm{est}} \, w_c h_c,
\end{equation}
where $w_c$ and $h_c$ are the width and height of connection $c$, respectively. In 3D-VASE, $\eta_s = 2 / \sqrt{2} = \sqrt{2}$ is used in the implementation. We then substitute $V_{\mathrm{est}}$ for the actual routed volume in Eq.~\ref{eq:tv_mismatch}, yielding the placement-stage validation rule for any converging paths $A$ and $B$:
\begin{equation}
\left|V_{\mathrm{est}}(A) \, Q_B - V_{\mathrm{est}}(B) \, Q_A\right| \le \epsilon.
\end{equation}
By embedding this inequality into the MIP model as a hard constraint, 3D-VASE couples placement with downstream
routability. This aggressively prunes invalid coordinate assignments with extreme distance disparities, guaranteeing
sufficient geometric space for subsequent 3D volumetric routing.

\subsubsection{\textbf{Dynamic Volume Validation in the Routing Stage}}
Merely reaching the target coordinates during $A^*$ search is insufficient to guarantee volumetric precision. Therefore,
we modify the routing termination logic from a purely geometric condition to a dual-validation rule that requires both
coordinate matching and volume matching:
\begin{equation}
\mathcal{C}_{\mathrm{stop}}: (\mathbf{p}_n = \mathbf{p}_e) \land \left(\left|V_{\mathrm{acc}}(n) - V_{\mathrm{req}}\right| \le \delta\right),
\end{equation}
where $\mathbf{p}_n$ and $\mathbf{p}_e$ denote the spatial coordinates of the current node and the target node, $V_{\mathrm{acc}}(n)$ is the accumulated physical channel volume from the source to node $n$, $V_{\mathrm{req}}$ is the required target volume for the routed connection, and $\delta$ is the admissible volume tolerance. Crucially, this Boolean condition dictates the detour generation mechanism. If a search branch arrives at the target $\mathbf{p}_e$ prematurely, namely when $V_{\mathrm{acc}}(n) < V_{\mathrm{req}} - \delta$, the algorithm strictly rejects this state. Deprived of the shortest-path solution, the $A^*$ priority queue is mathematically compelled to evaluate sub-optimal spatial branches, pushing the search frontier into adjacent unoccupied grids. This intentional rejection organically forces the channel to absorb the required hold-up volume before it is physically terminated at the target module.

\begin{table*}[t]
\centering
\small
\setlength{\tabcolsep}{3.5pt}
\caption{Comparison of synthesis results between 3D-VASE and 3M-DeSyn \cite{zhang20253m}.}
\label{tab:synthesis_results}
\resizebox{\textwidth}{!}{%
\begin{tabular}{l c c c c c c c c c c c c}
\toprule
\multirow{2}{*}{Test Case} & \multirow{2}{*}{V} & \multirow{2}{*}{E} & \multirow{2}{*}{T/V Const.} & \multirow{2}{*}{\#Layers} & \multicolumn{2}{c}{\#Vias} & \multicolumn{2}{c}{Channel Length $L$ ($\mu$m)} & \multicolumn{2}{c}{Chip Size ($\mu$m $\times$ $\mu$m)} & \multicolumn{2}{c}{Runtime (s)} \\
\cmidrule(lr){6-7} \cmidrule(lr){8-9} \cmidrule(lr){10-11} \cmidrule(lr){12-13}
& & & & & \textbf{3D-VASE} & 3M-DeSyn & \textbf{3D-VASE} & 3M-DeSyn & \textbf{3D-VASE} & 3M-DeSyn & \textbf{3D-VASE} & 3M-DeSyn \\
\midrule
Case 1 & 5 & 4 & N & 1 & 0 & 0 & 4200 & 8849 & $6850 \times 4200$ & $5900 \times 5465$ & 0.5 & 8.9 \\
Case 2 & 5 & 4 & Y & 1 & 0 & 0 & 15750 & 11341 & $10850 \times 7000$ & $9200 \times 6300$ & 2.0 & 32.3 \\
Case 3 & 7 & 6 & Y & 2 & 2 & 4 & 14300 & 14970 & $8200 \times 5200$ & $10100 \times 7300$ & 8.4 & 125.0 \\
Case 4 & 18 & 17 & N & 3 & 7 & - & 21300 & - & $13400 \times 7200$ & - & 30.3 & OOT \\
Case 5 & 27 & 26 & N & 1 & 0 & - & 82500 & - & $19450 \times 13850$ & - & 35.2 & OOT \\
Case 6 & 21 & 20 & Y & 2 & 6 & - & 38700 & - & $20900 \times 6500$ & - & 31.2 & OOT \\
Case 7 & 9 & 8 & Y & 3 & 5 & 8 & 14250 & 28900 & $10300 \times 5600$ & $11600 \times 8400$ & 13.0 & 410.8 \\
Case 8 & 9 & 8 & N & 1 & 0 & 0 & 8700 & 19300 & $16800 \times 5200$ & $15900 \times 8200$ & 2.1 & 178.5 \\
\bottomrule
\multicolumn{13}{p{0.97\textwidth}}{\small Notes: \textbf{V} = number of modules; \textbf{E} = number of connections; \textbf{T/V Const.} indicates whether timing and volumetric constraints are enabled (Y) or disabled (N); \textbf{OOT} = solver timeout at 3600 s; ``-'' indicates no valid baseline result within the timeout.}
\end{tabular}%
}
\end{table*}
\subsection{Hierarchical Conflict Resolution}

Dense layouts, especially single-layer designs without vertical escape paths, may exhaust BSG grid nodes as normal routing congestion and volume-induced detours compete for space. 3D-VASE therefore uses a two-level conflict strategy: local rip-up-and-reroute for isolated blockages and global placement feedback when the footprint itself becomes unroutable.
\subsubsection{\textbf{Local Rip-up and Reroute}}
A local deadlock occurs when the $A^*$ queue for channel $c_i$ is exhausted before reaching its target, indicating blockage by existing routes. 3D-VASE then removes one previously routed channel $c_j$ ($j<i$) with minimal expected disruption. Let $n_{fail}$ be the stalled grid node. The rip-up score is
\begin{equation}
Score_{ripup}(c_j) = \alpha \cdot Dist_{min}(n_{fail}, c_j) + \beta \cdot Length(c_j),
\end{equation}
where $Dist_{min}(n_{fail}, c_j)$ is the shortest Manhattan distance from $n_{fail}$ to path $c_j$, $Length(c_j)$ is its routed length, and $\alpha,\beta$ balance proximity and rerouting cost. The lowest-score channel is ripped up, its grid nodes are released, and both $c_i$ and the removed channel are returned to the routing queue.

\subsubsection{\textbf{Routability-Driven Global Feedback}}
If repeated rip-up fails and the counter exceeds $Threshold_{ripup}$, 3D-VASE treats the current placement as unroutable and restarts Stage 1 with relaxed geometry. The minimum module clearance is enlarged ($\rho \leftarrow \rho + \Delta\rho$), and an outer routing ring of width $w_{track}$ is reserved along the chip boundary. These changes slightly increase chip area but create additional internal and peripheral tracks, allowing the router to escape dense local congestion and converge under hard volumetric constraints.

\begin{figure}[t]
\centering
\includegraphics[width=0.85\columnwidth]{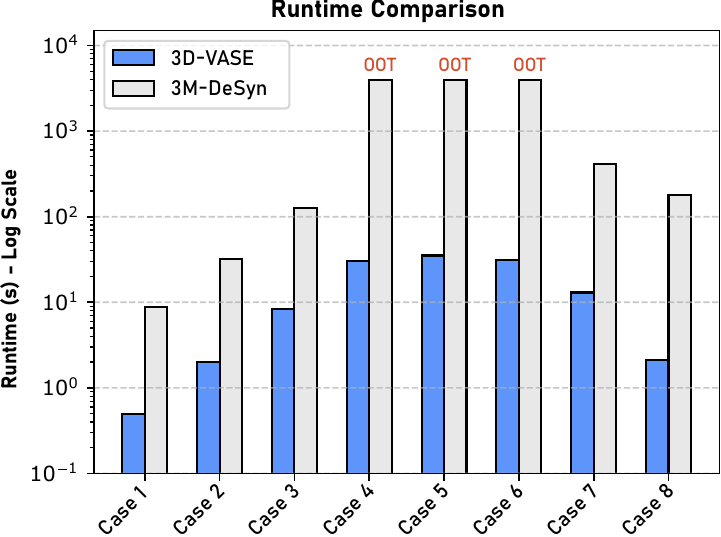}
% \vspace{-0.2cm}
\caption{Runtime comparison across benchmark cases. The runtime plot is presented on a logarithmic scale to highlight order-of-magnitude performance differences between 3D-VASE and 3M-DeSyn \cite{zhang20253m}.}
\Description{A runtime comparison chart across benchmark cases. The runtime axis is plotted on a logarithmic scale to emphasize large performance gaps between 3D-VASE and the baseline.}
% \vspace{-0.5cm}
\label{fig:evaluation_results}
\end{figure}

\section{Experimental Results}
\subsection{Experimental Setup and Baseline Method}
The core algorithms of 3D-VASE were implemented in Java, invoking the Gurobi 12.0.2 commercial solver to execute the Mixed-Integer Programming (MIP) computations during the global placement stage. All comparative experiments were executed on a workstation equipped with an Intel Core i9-13900H 2.6 GHz processor and 16 GB of physical memory.

Unlike conventional digital microfluidics or active-valve planar chips, 3D-printed microfluidics with passive flow control represents a highly specialized EDA domain. To the best of our knowledge, 3M-DeSyn is the most directly comparable synthesis framework for multi-layer 3D-printed microfluidics with timing and volumetric constraints. It represents the state-of-the-art approach in this specific domain. Therefore, we select 3M-DeSyn as our primary baseline to provide a rigorous and challenging evaluation of 3D-VASE. Methodologically, 3M-DeSyn employs a single-step global Constraint Optimization Programming (COP) model, abstracting channels as continuous polylines with discrete orientation variables. It relies on a massive array of Big-M constraint equations to simultaneously solve coordinate assignment, cross-layer routing paths, and non-overlapping/fluidic constraints. To ensure absolute fairness, this baseline method was evaluated in the identical hardware and solver environment.

In the context of agile microfluidic hardware development, design space exploration heavily depends on rapid synthesis feedback. If a single physical synthesis iteration takes hours, the engineering utility of automated synthesis tools is largely lost. We therefore treat 1 hour (3600 seconds) as a practical upper bound for one synthesis run in iterative design use; any method that cannot return a solution within this window is considered Out of Time (OOT).

\begin{figure*}[t]
    \centering
    \includegraphics[width=\linewidth]{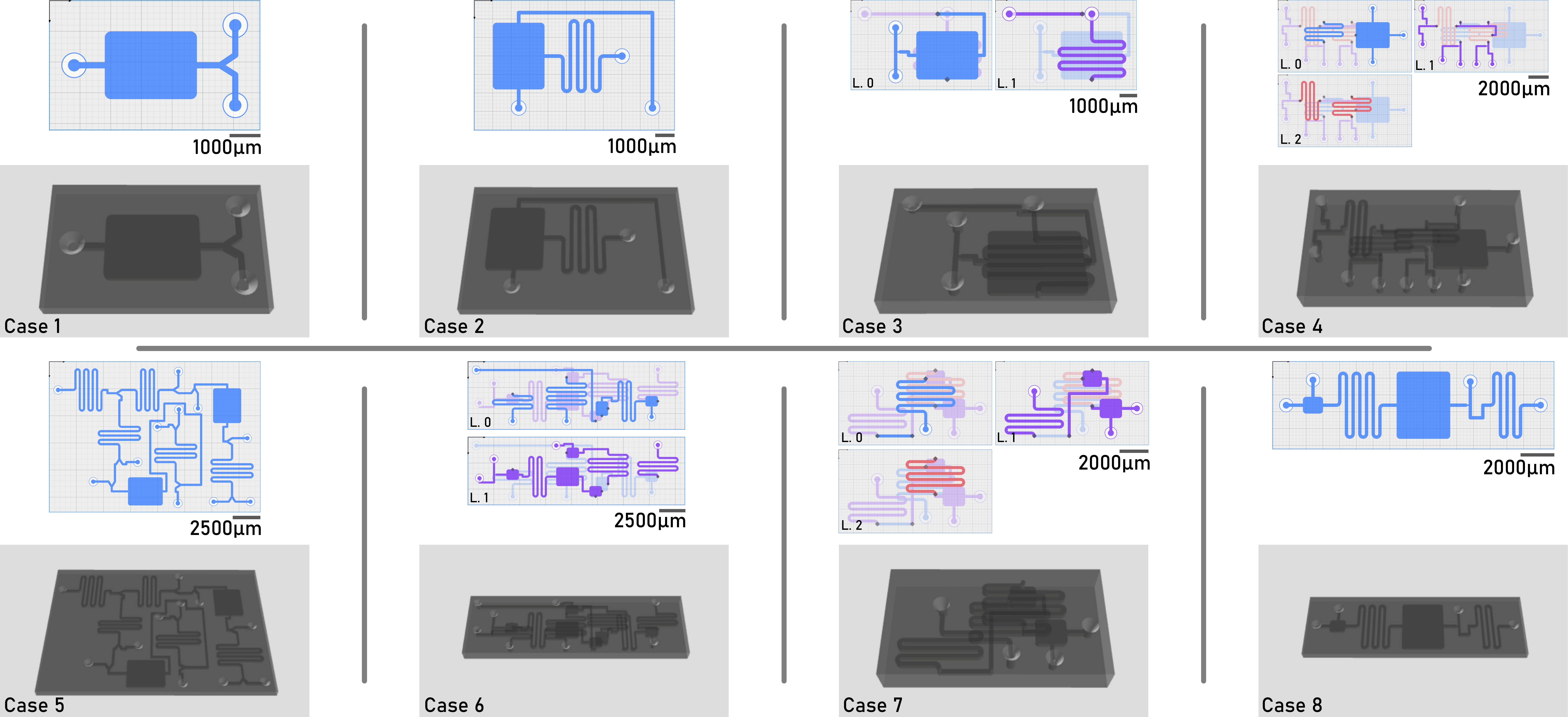}

    \caption{Synthesis results for representative microfluidic design test cases. For each case (1–8), the top row shows the generated 2D layout(s) with channel routing and component placement, while the bottom row presents the corresponding 3D-rendered model for fabrication. Multi-layer designs are visualized with color-coded layers. The results demonstrate scalability from simple single-channel layouts to complex, multi-layer integrated microfluidic networks.}
    \Description{For each benchmark case, the figure presents a 2D planar layout of the synthesized microfluidic routing result and the corresponding 3D-printable STL model derived from that layout.}
    \label{fig:testcases_results}
 \vspace{-0.3cm}
\end{figure*}

\subsection{Test Cases}
We use eight benchmarks with increasing scale and constraint complexity. Cases 1, 2, and 8 cover small to medium single-layer designs, with Case 2 adding volumetric constraints. Cases 3 and 7 represent multi-layer synchronization networks that require vertical transitions under timing and volumetric constraints. Case 4 stresses large 3D routing across three layers, Case 6 combines dense obstacle avoidance with volume-aware routing, and Case 5 is an extreme single-layer congestion case with 27 modules and 26 connections. Together, these cases evaluate scalability, via usage, volumetric detours, and fallback routability under limited routing resources.

\subsection{Performance and Quality Analysis}
Table~\ref{tab:synthesis_results} reports via counts, channel lengths, chip dimensions, and runtimes. Figure~\ref{fig:evaluation_results} visualizes selected comparative trends, while Figure~\ref{fig:testcases_results} shows the synthesized layouts and corresponding 3D-printable models.

\subsubsection{\textbf{Scalability and Runtime}}
As demonstrated in Table~\ref{tab:synthesis_results}, 3D-VASE improves runtime across all netlist scales. On small cases, it finishes in seconds, while 3M-DeSyn is substantially slower (e.g., 2.1 s versus 178.5 s in Case 8). More importantly, on the large cases 4--6, the monolithic baseline times out because it jointly reasons about placement, routing, collision constraints, and fluidic constraints. As net size grows or timing and volumetric requirements force longer detours, the number of candidate segments and Big-M disjunctions increases rapidly, causing branch-and-bound to spend most effort on infeasible combinations.

In contrast, 3D-VASE still finds valid layouts within 35.2 s, confirming that separating placement from BSG-based routing avoids the major combinatorial bottleneck. The global feedback loop adds little overhead in practice: most cases require no restart, while the hardest cases converge after at most two global restarts. This behavior indicates that local rip-up handles most congestion, and global expansion mainly acts as a lightweight safeguard for extreme footprints rather than as a costly outer optimization loop.

\subsubsection{\textbf{Resource Efficiency and Routing Topology}}
Layout quality also improves in most cases. The reduced Stage-1 search space allows the placement model to find compact chip bounding boxes before detailed routing begins, whereas the baseline must optimize placement, routing, and fluidic variables simultaneously. In Cases 3 and 7, chip area is reduced by about $42\%$ and $41\%$, respectively. The via penalty reduces cross-layer transitions: Case 3 uses 2 vias instead of 4, and Case 7 reduces vias from 8 to 5 while shortening channel length from 28900 $\mu$m to 14250 $\mu$m. Reducing vias is also beneficial for fabrication because each vertical transition can introduce additional dead volume and alignment-sensitive geometry.

Case 2 illustrates the main trade-off of the decoupled flow. Once placement is fixed, detailed routing may require extra boundary detours to satisfy timing/volume constraints and preserve routability, producing a larger footprint than the baseline in this case. Nevertheless, across the full benchmark set, the decoupled flow provides a more scalable runtime-quality trade-off and avoids the timeout failures observed in the monolithic baseline.

\section{Conclusion And Future Work}
This paper presents 3D-VASE, a two-stage physical synthesis framework for multi-layer 3D-printed microfluidic chips. By decoupling MIP-based global placement from heuristic BSG routing, 3D-VASE avoids the scalability bottleneck of monolithic optimization while preserving timing and volumetric feasibility through dynamic volume validation and hierarchical conflict resolution.

Experiments against 3M-DeSyn show clear runtime gains across all tested scales and convert several hard instances from timeout to practical runtimes. Layout quality also improves in most cases through tighter chip footprints and more practical routing outcomes. These results lower the computational barrier to complex fluidic physical design and support faster design-space exploration. More broadly, the proposed decoupled strategy suggests that physically constrained microfluidic synthesis can be made scalable without discarding fluidic correctness. This is especially important for early-stage design iteration, where designers need rapid feedback before committing to fabrication. Our validation is algorithmic under ideal steady-state pumping; future physical experiments should further quantify the impact of pump non-idealities and manufacturing variation.

Future work will relax the constant cross-sectional-area assumption and explore joint optimization of routing geometry and channel dimensions. Controlled channel-width variation could shorten paths and reduce chip footprint, but it requires localized fluidic constraints to maintain reliable flow velocity and avoid adverse effects such as sedimentation or clogging.

{
\begin{acks}

This work is supported by the Deutsche Forschungsgemeinschaft
(DFG, German Research Foundation) – Project Number 515003344.

\end{acks}
% \vspace{0.3cm}
}

\bibliographystyle{ACM-Reference-Format}
\bibliography{main}
\end{document}